\documentclass[12pt]{iopart}
\usepackage{epsfig,harvard,amsbsy,bm}
\newcommand{\isum}%
{\mathop{\hbox{$\displaystyle\sum\kern-13.2pt\int\kern1.5pt$}}}

\begin{document}
\bibliographystyle{jphysicsB}
\baselineskip = 8mm

\title[Kramers-Henneberger calculations of multi-photon ionization]
{On the use of the Kramers-Henneberger Hamiltonian in
multi-photon ionization calculations.}

\author{I. A. Ivanov\footnote[1]{Corresponding author:
Igor.Ivanov@.anu.edu.au}\footnote[2]{On leave from the Institute of
Spectroscopy, Russian Academy of Sciences} and A. S. Kheifets}
\address{Research School of Physical Sciences and Engineering,
The Australian National University,
Canberra ACT 0200, Australia}

\date{\today}
\begin{abstract}
We employ the Kramers-Henneberger Hamiltonian for time-independent
calculations of multi-photon ionization of atoms with one and two
electrons.  As compared to the electromagnetic interaction in the
length and velocity gauges, the presently employed Kramers-Henneberger
gauge has an advantage of the dipole matrix elements for the free-free
electron transitions being finite and well-defined quantities. This
circumstance simplifies considerably the computations and allows to
obtain accurate results for the two-photon ionization of realistic
atomic systems.
\end{abstract}

\maketitle

\section{Introduction}
\label{S1}

Recent years have witnessed a rapid advancement of high-power
short-pulse laser techniques which make it possible to observe many
striking phenomena such as multi-photon ionization (MPI) and
above-threshold ionization (ATI). This progress in experimental
techniques has been accompanied by equally rapid development of
theoretical methods needed to describe adequately phenomena occurring
in strong laser fields. Representative reviews giving detailed account
of these theoretical advances can be found in \citeasnoun{PKK97},
\citeasnoun{LMZ98}, \citeasnoun{CT04} and \citeasnoun{Posthumus04}.

It is well-known that interaction of an atom and the electromagnetic
(EM) field can be described in various ways corresponding to different
choices of the gauge. In single-photon ionization calculations, it is
the length and velocity gauges that are used most commonly.  This
choice is quite natural since in the length and velocity gauges the
dipole matrix elements assume a very simple form.  If, however, one
wishes to compute probabilities of various multi-photon processes, a
difficulty immediately arises. The dipole matrix elements
corresponding to the free-free electron transitions are divergent.

There are various ways to circumvent this difficulty. For one-electron
systems this problem can be easily avoided since analytical
expressions for the Coulomb Green function are known. This fact has
been exploited in many papers
\cite{KL69,RAP69,ARN73,TL76,KM79,Kar71,Kar85,Kar03}.
Alternatively, one may avoid the divergency problems by reducing
summations and integrations over intermediate electron states to
solution of nonhomogeneous differential equations
\cite{ZK65,CT69,JT2001}.

For systems with more than one electron, where these techniques cannot
be implemented, other methods have been developed.  
The states belonging to the continuous spectrum of
the system can be represented 
by means of a suitable set of the square integrable ($L^2$)
functions \cite{Bach93,Ven96}.  The system is quantized
in a box of sufficiently large dimensions which gives a discretized
representation of the continuous spectrum.  This approach,
supplemented by the B-splines technique, allowed to obtain a set of
accurate MPI and ATI cross-sections for many-electron systems such as
He \cite{Ven96,SL99,NL01} and Be \cite{Bach93}.  Another technique
which was applied successfully to study MPI in many-electron systems
uses a regularization procedure for the free-free matrix-elements
\cite{mn1,mn2,Kor97}.

In the present paper we describe an alternative method allowing to
compute amplitudes of the MPI and ATI processes in many-electron
systems in a more direct way. The proposed method is based on the
so-called Kramers-Henneberger (KH) form of the interaction
Hamiltonian \cite{kh1,kh2}. 
The KH representation of the system ``atom plus electromagnetic
field'' is often used when one studies a temporal evolution of atomic
systems subjected to a pulse of EM radiation \cite{RB,VK}. The KH
representation also enables simple asymptotic boundary conditions used
in the external  region of the $R$-matrix Floquet theory
\cite{BFJ91}. 

In the present paper we shall be interested in another aspect of the KH
representation, namely the advantages its use may offer in the
perturbative computations of MPI rates. To our knowledge, the KH
description of the EM radiation interaction with atoms has not been
used in this context. As we shall see, in the perturbative
calculations the KH representation offers one important
advantage. In contrast to the length and velocity gauges, in the KH
formulation all the dipole matrix elements are finite and
well-defined.
For the laser fields of not very large intensities (below
$10^{13}$~Wcm$^{-2}$) the perturbation theory (PT)
provides quite an adequate description of the MPI process and allows
to achieve numerically accurate results with much less computational
labor. We shall consider below two-photon ionization processes in the
hydrogen and helium atoms.
The  highly accurate perturbative results available in the literature
allow us to evaluate directly
the accuracy of the method. We also discuss some subtle
numerical aspects of  application of the KH Hamiltonian
in perturbative calculations.

The use of the KH Hamiltonian in time-independent MPI calcualtions is
not restricted to preturbative regime. In our recent paper
\cite{nous1} we demonstrated  utility of this method  in a 
non-perturbative calculation of the MPI rates in atomic
hydrogen.

In the following sections, we briefly recall main theoretical aspects
of the KH representation, describe the computation of the dipole
matrix elements and apply the technique to two-photon ionization of
atomic hydrogen and helium.

\section{Theory}
\label{SH}

A starting point for the description of the interaction of an atom and
a monochromatic EM field is the minimal coupling Hamiltonian:
\begin{equation}
\hat H = \hat H_{\rm atom}+ \hat H_{\rm int} \ .
\label{minc}
\end{equation}
Here $\hat H_{\rm atom}$ has the usual meaning of the atomic
Hamiltonian:
\begin{equation}
\hat H_{\rm atom}=\sum\limits_{i=1}^{N}{{\bm p}_i^2\over 2}
-\sum\limits_{i=1}^{N}{Z\over r_i}
+\sum\limits_{i,j=1,i>j}^{N}{1\over r_{ij}}
\label{hams}
\end{equation}
The atomic Hamiltonian is taken in a non-relativistic form with $Z$
being the nucleus charge.
The part of the Hamiltonian $ \hat H_{\rm int}$ which describes the
interaction of the atom and the EM field can be written as (see
e.g. \citeasnoun{Sobelman72})
\begin{equation}
\hat H_{\rm int}=-{1\over c}\sum\limits_{i=1}^{N}
\left(\hat {\bm A} \cdot \hat {\bm p}_i-
{\hat{\bm A^2}\over 2 c^2}\right),
\label{hint}
\end{equation}
where $\hat {\bm p}$ is the momentum operator, $\hat {\bm A}$ is a
vector potential, summation runs over all atomic electrons. In the
following we shall assume that the dipole approximation is valid so that
the vector potential does not depend on atomic coordinates.
Performing a suitable canonical transformation of \Eref{hint}, one can
obtain various forms of the interaction Hamiltonian.  The KH Hamilton
is obtained by the canonical transformation
$\displaystyle \hat H_{\rm
KH}= e^{i\hat T}\hat H_{\rm min}e^{-i\hat T}-{\partial \hat T\over \partial
t}$
generated by the operator:
\begin{equation}
\hat T=-{1\over c}\int\limits_0^t
\sum\limits_{i=1}^N{\bm A}(\tau){\bm p_i}\ d\tau+
       {1\over 2c^2}\int\limits_0^t {\bm A}^2(\tau)\ d\tau \ ,
\label{khf}
\end{equation}
Expressed in quantum-mechanical terms (as far as description of the EM
field is concerned), this transformation is also known as the
Pauli-Fierz canonical transformation \cite{PF}.  We shall not
distinguish between these two versions of the transformation as the
final results are identical.

Under the transformation (\ref{khf}) the minimal-coupling
Hamiltonian (\ref{hams}) becomes:

\begin{equation}
\hat H_{\rm KH}=\hat H_{\rm atom}+\hat H_{\rm int}^{\rm KH} \ ,
\label{hams1}
\end{equation}
where $\hat H_{\rm atom}$ retains the same form
as the Hamiltonian (\ref{hams}) while the
interaction Hamiltonian becomes:
\begin{equation}
\hat H_{\rm int}^{\rm KH}=\sum\limits_{i=1}^N
\left({Z\over r_i}-{Z\over |\bm r_i+\hat{\bm \alpha}|}\right),
\label{hint1}
\end{equation}
We shall be interested in the case of a linearly polarized
monochromatic EM field.  In this case
$
\hat{\bm
\alpha}=\hat{\bm F}/ \omega^2$
where $\hat{\bm F}$ is the operator of
the electric field intensity, $\omega$ is the photon energy.
\footnote[3]{We  use the atomic system of units in which $\hbar=e=m=1$ }.
If $\hat{\bm \alpha}$ can be considered as a small quantity,
the leading term of expansion of (\ref{hint1}) reproduces
the well-known $Z{\bm r}/r^3$ form of the interaction
Hamiltonian in the acceleration gauge. This form is often used
in the first-order perturbation calculations. We, however, are
interested in higher order effects and must
generally retain complete form of the Hamiltonian (\ref{hint1}).

To build the perturbation theory expansion, treating operator
(\ref{hint1}) as a perturbation, we need a formula for the matrix
elements of this operator sandwiched between the states describing the
noninteracting atom and the EM field.  It is convenient for our purposes
to use the notation $|a, m\rangle$ for these states where $a$ stands
for a set of quantum numbers describing the atom and $m$ denotes a
number of laser photons in a given mode.
Such a formula can be obtained from the known
matrix elements of the quantized vector potential operator
\cite{nous1}. A simpler derivation, relying on the
correspondence between the quantum and classical description of the EM
field, is given in the Appendix I.

Obtained in either way, the final formula reads:
\begin{equation}
\left\langle a, n+p\left|\hat H_{\rm int}^{\rm KH}\right|
b, n\right\rangle=
{1\over \pi} \sum\limits_{i=1}^N
\int\limits_0^{\pi}
\cos{p\theta}
\left\langle a\left|\ {Z\over r_i}-{Z\over |\bm r_i+
{\bm F}\cos{\theta}/ \omega^2|}\right|b\right\rangle
\ d\theta
\label{ma6}
\end{equation}
Here $\bm F$ is already a classical vector and not an operator. Its
magnitude is related to the number of photons via the
flux conservation relation $F^2/8\pi=n\omega$, and
it is directed along the polarization vector of the incoming
photons. In the following, we shall take this direction
as the $z$-axis.

To amend \Eref{ma6} to a form suitable
for practical computations, we use the well-known expansion:

\begin{equation}
{1\over \Big|\bm r+\bm F\cos\theta/ \omega^2\Big|}=
\sum\limits_{k=0} \sqrt{4\pi\over 2k+1}{r_<^k\over r_>^{k+1}}
[-{\rm sign}(\cos{\theta})]^k \ Y_{k0}({\bm r}),
\label{ma7}
\end{equation}
where $r_<$ ($r_>$) is the smaller (greater) of $r$ and $F\cos\theta/
\omega^2$.  
\Eref{ma7} allows separation of the  radial and angular
variables.
Angular parts are evaluated analytically using integrals of
products of several spherical functions \cite{Varshalovich}:
\begin{eqnarray}
\int Y_{l_1m_1}({\bm \Omega})
Y_{l_2m_2}({\bm \Omega})Y_{l_3m_3}({\bm \Omega})\ d{\bm \Omega}
=\nonumber \\
\sqrt{
(2l_1+1)(2l_2+1)(2l_3+1)\over 4\pi}
\left(\matrix{l_1 & l_2 & l_3 \cr 0 & 0 & 0}\right)
\left(\matrix{l_1 & l_2 & l_3 \cr m_1 & m_2 & m_3}\right)
\label{var}
\end{eqnarray}
Equation (\ref{var}) is written for the case when the atomic subsystem
contains one electron (hydrogen). Generalization for the case of
many-electron systems is a simple exercise in angular momentum
algebra.

When performing perturbation calculations, we are usually interested
in keeping track of field dependencies of the matrix elements. Suppose
we study a process for which, in the leading order, the amplitude is
proportional to $k$-th power of electric field strength 
$M\propto F^k$,
with some integer $k$. Then we would like to retain in the
perturbation theory expressions only the terms which  give rise to
such dependence in the limit of small $F$.  If the length or velocity
forms for the atom-EM field interaction is used, such a count of
powers of $F$ is trivial, following from the well-known selection rules
for the matrix elements. In the KH representation, there are no exact
selection rules. Nevertheless, the count of powers of the electric
field strength is still possible. Consider, for example, the case of a
two-photon ionization of a one-electron atom, which is the second-order
process.  Suppose, we are interested in the ionization from the state
$a$ with a given orbital momentum $l$.  Then it is easy to see from
\Eref{ma7} that the following asymptotics holds for $F\to 0$:
\begin{equation}
\label{asymptotics}
\langle al\ m|\hat H_{\rm int}^{\rm KH}|bl\ m\pm
2\rangle \propto F^2
\ \ \ \ , \ \ \
\langle al\ m|\hat H_{\rm int}^{\rm KH}|bl\pm
1\ m\pm 1\rangle \propto F \ .
\end{equation}
Here $a$ and $b$ stand for the set of all 
atomic quantum numbers except the
angular momentum, the integer $m$ refers to a total number of laser
photons.  It is easy to see from \Eref{ma7} that the coefficient of
proportionality in the second matrix element of \eref{asymptotics} is
just a matrix element of the operator $Z{\bm r}/r^3$ which is commonly
used in the first order calculations in the acceleration gauge.

Thus, the leading terms of the amplitude of the two-photon ionization
can be written as:
\begin{eqnarray}
M(al\ m\to bl\ m-2)=
\langle al\ m|\hat H_{\rm int}^{\rm KH}|bl\ m-2\rangle \nonumber \\
+
\isum\limits_{cl'\ l'=l\pm 1}
{\langle al\ m|\hat H_{\rm int}^{\rm KH}|cl'\ m-1\rangle
\langle cl'\ m-1|\hat H_{\rm int}^{\rm KH}|bl\ m-2\rangle
\over E_{al}-E_{cl'}+i\epsilon}
\label{amp1}
\\
M(al\ m\to bl+2\ m-2)=
\langle al\ m|\hat H_{\rm int}^{\rm KH}|bl+2\ m-2\rangle \nonumber \\
+
\isum\limits_{cl'\ l'=l+1}
{\langle al\ m|\hat H_{\rm int}^{\rm KH}|cl'\ m-1\rangle
\langle cl'\ m-1|\hat H_{\rm int}^{\rm KH}|bl+2\ m-2\rangle
\over E_{al}-E_{cl'}+i\epsilon}
\label{amp2}
\\
M(al\ m\to bl-2\ m-2)=
\langle al\ m|\hat H_{\rm int}^{\rm KH}|bl-2\ m-2\rangle \nonumber \\
+
\isum\limits_{cl'\ l'=l-1}
{\langle al\ m|\hat H_{\rm int}^{\rm KH}|cl'\ m-1\rangle
\langle cl'\ m-1|\hat H_{\rm int}^{\rm KH}|bl-2\ m-2\rangle
\over E_{al}-E_{cl'}+i\epsilon}
\label{amp3}
\end{eqnarray}
Here the symbol $\isum$ indicates the sum over the discrete spectrum
and integration over continuum of the intermediate states. Without
sacrifice of accuracy, in these sums we may use the operator $Z{\bm
r}/r^3$ instead of the complete form of the operator $\hat H_{\rm
int}^{\rm KH}$.

Once the amplitudes \eref{amp1} -- \eref{amp3} are computed, the
generalized partial cross-section of the two-photon ionization from
the initial state $a,l$ to a  final channel $b,l'$ is given by
(c.f. \citeasnoun{Bach93})
\begin{equation}
\sigma(al\to bl')
=2^7\pi^3\alpha^2 a_0^4\tau_0\omega^2
\lim\limits_{F\to 0} {|M(al\to bl')|^2\over F^4 k} \
[{\rm cm}^{4}{\rm s}^{-1}]
\label{crs}
\end{equation}
Here $\alpha$, $a_0$, $\tau_0$ are the fine structure constant, the
atomic unit length in cm and the atomic unit time in seconds.  $F$ is
the EM field strength and $\omega$ is the photon energy, both
expressed in the atomic units. The one-electron continuum wave
functions used to calculate the ionization amplitudes are normalized
on the momentum scale.  The generalized cross-section (\ref{crs}) is
related to the ionization rate
\begin{equation}
\Gamma(al\to bl')={\sigma(al\to bl')\times 10^{12}\lambda\ {\rm Ryd}\over
13.605\times 1.60219} \
[{\rm W}^{-1}{\rm cm}^4]
\ ,
\label{rate}
\end{equation}
where $\lambda$ is wavelength (in nm) and ${\rm Ryd}=109677$
cm$^{-1}$ is the Rydberg constant \cite{Kar03}.

It has to be noted that the KH transformation modifies, in general,
the atomic states after the field is switched off at a sufficiently
large time \cite{VK,RB}. However, it is shown in Appendix II that this
does not affect the cross-sections or ionization rates and can only be
noticed in the fine details of the temporal evolution of the system

\section{Numerical computations}
\label{SH3}

Certain amount of care has to be exercised when the amplitudes
\eref{amp1} -- \eref{amp3} are computed numerically.  Consider matrix
elements in the sums over intermediate states in these
expressions. Separating radial and angular variables with the help of
equations \eref{ma6}, \eref{ma7}, and taking the form $Z{\bm r}/r^3$
for the interaction Hamiltonian we obtain integrals of the sort
$I=\int R_{El}(r)R_{kl'}(r)\ dr$. Here $R$'s are radial electron wave
functions behaving as $r^l$ near the origin $r=0$. Function $R_{El}$
describes either an initial or final state of the process, $R_{kl'}$
is the radial wave function of the intermediate state belonging to the
continuous spectrum which, for large momenta, behaves as
$R_{kl'}(r)\propto \sin(kr+\delta)/r$.  The $k$-dependence of the
integral is crucially dependent on the orbital momentum $l$. 
If $l>0$ the integral $I$ can be approximated for large $k$ 
as $\int R_{El}(r) \sin(kr+\delta)/r$ (we omit unimportant normalization 
factors).  
Consider the function $P(r)=R_{El}/r$.
If $l>0$, $\int\limits_0^{\infty} |P(r)| dr$ is finite, hence the
Riemann-Lebesgues lemma is applicable and the integral $I$ will decay
for $k\to\infty$. If $l=0$, the replacement of the $R_{kl'}$
by its asymptotic expression is not legitimate (we would obtain a
divergent integral). More careful study shows that if $l=0$
than $I\to {\rm const}$ when
$k\to\infty$.  The integrals over momenta of intermediate
states in the formulas \eref{amp1} -- \eref{amp3} still converge in
this case due to the energy denominator but much more slowly than in
the case of $l>0$. This means that to achieve a good numerical accuracy
for the amplitudes with $l=0$, one must take into account an
asymptotic tail of the integrand for $k\to\infty$.

In practice, this does not pose serious difficulties.  One has only to
determine (either analytically or numerically) the constant in the
expression $I\to {\rm const}$ for $k$ large enough so that this
asymptotic law holds well, and then add the corresponding contribution
to the integrals in  formulas \eref{amp1} -- \eref{amp3}.  For $l>0$
one need not worry about the asymptotic tails of the integrand as
integrals $I$ decay quite fast with $k$.

For the presently considered targets (H and He), the initial state has
an $S$ orbital character leading to the two final channels $S$ and $D$
which correspond to amplitudes \eref{amp1} and \eref{amp2}. As
explained above, we may retain in these amplitudes only the $P$
intermediate states. Both the discrete and continuous intermediate
states were taken into account.  The continuous spectrum integration
follows closely prescriptions given in \citeasnoun{B94} and
\citeasnoun{nous1}. The interval of momenta $(0,q_{\max})$ is
divided into several subintervals.  For the photon energies above the
ionization threshold, a pole is present in the momentum integral. To
carry out  the integration around the pole accurately, the first two
subintervals are chosen to be $(0,k_{\rm pole})$ and $(k_{\rm
pole},2k_{\rm pole})$ with a typical number of 20 momentum points in
each subinterval.  Then the delta-function singularity is isolated and
the remaining principle value integral is evaluated by a modified
Gaussian rule \cite{B94}. The remaining part of the momentum integral
is divided as follows: $(2k_{\rm pole },4)$ (20 integration points),
$(4,10)$ (20 points) and $(10,q_{\max})$ (20 points).  These
intervals are pole-free and the integration is performed by using a
Gauss quadrature rule.  The fairly large value of $q_{\max}$ is
chosen to take care of a slow decay of the integrand in the
$S$-channel. The asymptotic tail $(q_{\max},\infty)$ is calculated
analytically.

It is worth to be noted that the first order matrix elements
in the amplitudes \eref{amp1}, \eref{amp2} are roughly of the same
magnitude as the second order terms and, sometime, of the opposite
sign. So their inclusion is essential.

Once the amplitudes are computed the partial cross-sections and rates
can be determined via Eqs.~(\ref{crs}), (\ref{rate}).  Total
cross-sections and rates corresponding to linear and circular
polarization of the EM field can then be determined. If inital
state of atomic system is an $S$-state, than: 
$\displaystyle \Gamma^l(ns)=\Gamma(ns\to
ks)+\Gamma(ns\to kd)$ (linear polarization) and $\displaystyle
\Gamma^c(ns)={3\over 2}\Gamma(ns\to kd)$ (circular polarization)
\cite{JT2001}.

All calculations reported below were performed for the EM field
strength of $F=0.03$ a.u. This field strength is small enough, so that
determination of the cross sections via Eq.(\ref{crs}) for small but
finite $F$ will be, as we shall see, quite accurate.

\section{Results}
\subsection{Hydrogen}

In the case of hydrogen, all the one-electron wave functions are know
analytically. These allowed us to include 15 discrete intermediate $P$
states and carry out integration up to $q_{\max}=70$~a.u. in the
second-order amplitudes.

\begin{table}[h]
\caption{\label{tab1}
Ionization rates (in units of W$^{-1}$cm$^4$) for the two-photon
ionization of atomic hydrogen in the ground state by linearly
$\Gamma^l$ and circularly $\Gamma^c$ polarized light. Numbers in
parenthesis indicate powers of 10.  The literature values are denoted
as JT01 \cite{JT2001} and KM03
\cite{Kar03} }
\bigskip

\small
\hspace*{1cm}
\begin{tabular}{l| lll | lll}
\br
$\lambda$
& \multicolumn{3}{|c|}{$\Gamma^l$}
& \multicolumn{3}{c}{$\Gamma^c$}\\
nm   &JT01&KM03& Present
     &JT01&KM03& Present \\
\mr
  20    &3.02(-38)& 3.01(-38)  &3.02(-38)&2.44(-38)& 2.43(-38)&2.44(-38)  \\
  40    &2.15(-36)& 2.14(-36)  &2.15(-36)&2.03(-36)& 2.02(-36)&2.03(-36)  \\
  60    &2.62(-35)& 2.61(-35)  &2.62(-35)&2.79(-35)& 2.78(-35)&2.79(-35)  \\
  80    &1.58(-34)& 1.57(-34)  &1.59(-34)&1.85(-34)& 1.84(-34)&1.86(-34)  \\
\br
\end{tabular}
\normalsize
\end{table}

In \Tref{tab1}, we present the rates of the two-photon ionization of
the ground state hydrogen atom with the linear and circular polarized
light and compare them with the latest literature values
\cite{JT2001,Kar03}.  The three sets of data  are virtually identical
for all the photon energies considered. The cited literature values
have been obtained analytically which is possible for hydrogen. They
can be, therefore, considered as "exact''. The comparison in the Table
indicates that the use of the KH Hamiltonian allowed us to achieve
comparable level of accuracy with little effort.

\subsection{Helium}

We consider now two-photon ionization from the ground state of
helium. Due to possible two-electron excitations, complete treatment
of this problem is much more complicated than for hydrogen.
However, for the photon energies below the threshold of the $N=2$
excitations, we can use a frozen core approximation and restrict
ourselves with only one active electron. In this approximation, the
problem is effectively reduced to a hydrogen-like calculation with
numerical Hartree-Fock wave functions.

We describe the helium atom as follows. For the ground state we use
the self-consistent Hartree-Fock approximation \cite{CCR76}.  The
ground state is thus represented as a product of the two $1s$
orbitals. For the excited states, both discrete and continuous, we use
the frozen core Hartree-Fock approximation \cite{CCR79}. These states
are thus represented as properly symmetrized products of the $1s$ core
orbital and an orbital describing excited electron either in the bound
state or the continuum.

Because of this frozen-core approximation, the calculation for helium
proceeds in almost exactly the same way as for hydrogen. The only
difference is that due to the equivalence of the $1s^2$ electrons an
additional factor of 2 arises in the formula for the cross-section.
As for hydrogen, we may retain only $P$ states in the sums over the
intermediate states. We have retained 7 $P$-states corresponding to
the excitations $1snp$ with $n=2-8$. Integration over the continuous
spectrum was performed as for hydrogen except for the value of
$q_{\max}=30$~a.u.  All we said above concerning importance of the
correct account of the asymptotic tail in the momentum integral for
the $S$-wave, applies for the case of helium as well.

Our results are presented in the Figure in comparison with other
calculations \cite{SL99,NL01} which use a considerably more accurate
representation of the helium atom.  Despite a rather crude character
of the frozen-core approximation, in the region of the photon energies
considered we achieve quite a satisfactory agreement with the
literature values.  We could not, of course, extend our calculation
into the region of larger photon energies since there the processes of
core-excitations become essential.

\begin{figure}[h]
\epsfxsize=7cm
\epsffile{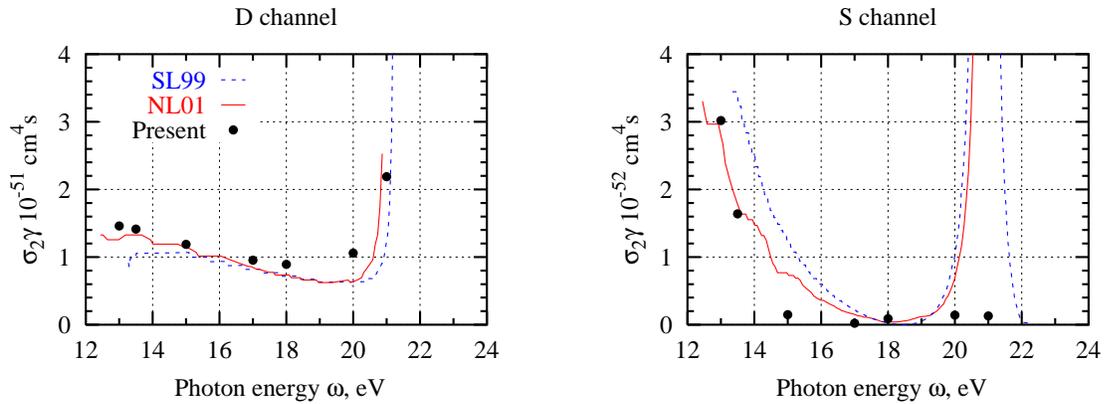}

\caption{\label{fig1}
Cross section of the two-photon ionization from the ground state of
helium  Comparison is made with literature values marked as SL99
\cite{SL99} and NL01 \cite{NL01}}
\end{figure}

\section{Conclusion}
\label{S5}

We have shown that the KH description of the atom-EM radiation
interaction can be used efficiently in calculations of MPI and ATI
processes in realistic atomic systems.  The fact that the dipole
matrix elements between continuous electron states are finite and
well-defined quantities makes the calculation relatively simple, both
numerically and conceptually.

For hydrogen, the present results agree completely with those obtained
in \citeasnoun{JT2001} and \citeasnoun{Kar03}.  As the latter results
are analytical and can therefore be considered as virtually exact, we
may be confident that the present approach allows to achieve quite a
high accuracy.  Our method can also be applied to the systems with
more that one electron, as the comparison with the data for the
two-photon ionization of helium shows. Despite the fact that we used
rather a crude description of the field-free helium atom (we omitted
the core-excitation effects), we obtained good quantitative agreement
with the results of other authors who employed a more elaborate
representation for the helium atom. As was indicated above, the
accuracy of our description of MPI of helium can be further improved
without any problems of conceptual character. All we have to do is to
``thaw'' the core and to allow the two-electron excitations.  This can
be done, for example, with the use of the convergent close-coupling
(CCC) method
\cite{B94} which is known to provide good description of a
complete set of two-electron states, both discrete and continuous.
Such calculation will be reported elsewhere.

\section{Appendix I}
\label{app}

We give below a derivation of \Eref{ma6} for the matrix elements of
the KH interaction Hamiltonian based on the well-known correspondence
between the classical Floquet and the quantum-mechanical descriptions
of the atom-EM field interaction \cite{Sh65}.

In the classical picture, the KH Hamiltonian has the form
\eref{hint1} with ${\bm \alpha} ={\bm F}\cos{\omega t}/ \omega^2$
where ${\bm F}$ is a classical amplitude of the EM field.
With this expression being a periodic function of time, the
Shcr\"odinger equation has a set of solutions (the Floquet anzats)
which allows the following Floquet-Fourier expansion:
\begin{equation}
\Psi(t)=e^{-iEt}\sum\limits_{n=-\infty}^{+\infty} u_n e^{in\omega t},
\label{fla}
\end{equation}
where $E$ is the quasi-energy.
Expanding the time-periodic function $\hat H_{\rm int}^{\rm KH}$ as a
Fourier series:
\begin{equation}
\hat H_{\rm int}^{\rm KH}(t)=\sum_{n=0}^{\infty} V_n \cos{n\omega t},
\label{four}
\end{equation}
with
\begin{equation}
V_n
={2\omega\over \pi} \int\limits_0^{\pi\over\omega}
\hat H_{\rm int}^{\rm KH}(wt)\cos{n\omega t}\ dt,
\label{four1}
\end{equation}
and equating coefficients
with $e^{-im\omega t}$, one obtains a set of equations for
the Fourier amplitudes $u_n$:
\begin{equation}
\left(E-n\omega-\hat H_{\rm atom}\right) u_n=
\sum\limits_{k,m,k\ge 0\atop m-k=n} {V_k\over 2} u_m+
\sum\limits_{k,m,k\ge 0\atop m+k=n} {V_k\over 2} u_m
\label{four2}
\end{equation}
In the quantum-mechanical description, the coefficients with the
amplitudes $u_m$ on the r.h.s of \Eref{four2} are nothing but the
matrix elements $\langle n|\hat H_{\rm int}^{\rm KH}|m\rangle$ taken
between the states with $n$ and $m$ photons \cite{Sh65}.  This
correspondence holds if we neglect all spontaneous processes and
retain only laser photons.
Since in Eq.(\ref{four2}) summation index $k\ge 0$, for given
$n$,$m$, $n\neq m$, the
two terms on the r.h.s of this equation can be
combined to give rise to $\displaystyle {V_{|n-m|}\over 2}$.
This gives immediately the
formula \eref{ma6} for the matrix elements of the operator $\hat
H_{\rm int}^{\rm KH}$. In our earlier paper \cite{nous1}, we obtained
this formula directly using quantized form of the electric field
operator.

\section{Appendix II}

A comment has to be made on the applicability in the present case 
of the Fermi golden rule which is used to derive Eq.(\ref{crs}). 
Below, all the
discussion uses only classical terms for the description of the EM
field which leads to the same results as the full quantum-mechanical
treatment but is somewhat simpler.

Physically, transformation to the KH frame is equivalent to
transformation to the non-inertial frame oscillating with the
electron. Let us suppose that an initially field-free atom is in some
state $\Psi_0$. At the moment $t=0$ the interaction of the atom and EM
field is switched on and it is switched off at the moment $t=t_1$ with
some $t_1$ large enough so that all the transient processes are
negligible. With the help of  
formulas \eref{amp1} -- \eref{amp3} we can obtain (for
sufficiently large $t_1$) perturbative solution of 
the time-dependent Schr\"odinger
equation in the KH-frame.  As a result, we obtain at the moment $t_1$ a
vector in the KH frame:
\begin{equation}
\Psi_{\rm KH}(t_1)=U^{\rm KH}(0,t_1)\Psi_0,
\label{psi}
\end{equation}
where $U^{\rm KH}(0,t_1)$ is the evolution operator which we
constructed in the KH frame to a given order of the perturbation 
theory. 

To determine rates of various processes, we must now find overlaps of
the vector thus obtained with various states of the field-free
atom. The latter, however, are generally different from the
eigenstates of $H_0$, the original field-free atomic Hamiltonian. As
formula \eref{khf} shows, they are connected to the eigenstates of
$H_0$ by means of a unitary transformation generated by the operator
$\hat T$ in Eq.~\eref{khf}. The fact that final states are generally
altered by this operator may play an important role in calculations of
time-evolution of an atomic system under the action of an impulse of
EM radiation
\cite{VK,RB}.

The problem we are considering here differs from the problems of
determining the time-evolution under the action of an impulse of EM
radiation in one important aspect. We are at liberty to switch off
interaction at any time (provided it is large enough). Operator $\hat
T$ in Eq.(\ref{khf}) contains two terms. The one proportional to the
square of the vector potential is unimportant (it is a pure phase
factor in the dipole approximation which we use). It is the term
linear in ${\bm A}$ that is responsible for the alteration of the
final states. We may use, however, periodicity of the vector potential
and choose the moment $t_1$ so that this term disappears. The final
states then remain unaltered and the validity of the Fermi golden
rule is restored.

\section{Acknowledgements}

The authors acknowledge support of the Australian Research Council in
the form of  Discovery grant DP0451211. Facilities of the Australian
Partnership for Advanced Computing (APAC) were used in this project.

\newpage


\end{document}